# Ferrimagnetism and spin canting of $Zn^{57}Fe_2O_4$ nanoparticles embedded in ZnO matrix

G. F. Goya[†] and E. R. Leite‡

[†] *Instituto de Física, Universidade de São Paulo, CP 66318 – SP05389-970 São Paulo, Brazil*

[‡] *Centro Multidisciplinar de Desenvolvimento de Materiais Cerâmicos, Departamento de Química, Universidade Federal de São Carlos CP 676, 13560-905, São Carlos, SP, Brazil*

**SHORT TITLE:** Ferrimagnetism and spin canting of $Zn^{57}Fe_2O_4$ nanoparticles.

[†] E-mail address: goya@unizar.es

## *ABSTRACT*

The structural and magnetic properties of $ZnFe_2O_4$ nanoparticles embedded in a non-magnetic ZnO matrix are presented. X-ray diffractograms and Transmission Electron Microscopy (TEM) images showed that the resulting samples are composed of crystalline ferrite nanoparticles with average crystallite size $\langle D \rangle = 23.4\pm0.9$ nm, uniformly dispersed within the ZnO matrix. Magnetization data indicated a superparamagnetic-like behavior from room temperature down to $T_M \sim 20$ K, where a transition to a frozen state is observed. The M(H) curves displayed nearly zero coercive field down to $T_M$, where a sharp increase in the $H_C$ value is observed. The measured saturation magnetization $M_S$ values at 200 and 2 K were $M_S = 0.028(3)$ and 0.134(7) $\mu_B$/f.u. $ZnFe_2O_4$ respectively, showing the existence of small amounts of non compensated atomic moments. Mössbauer measurements at low temperatures confirmed the transition to a magnetically ordered state for $T < 25$ K, where two magnetically split sextets develop. Whereas these two sextets show strong overlap due to the similar hyperfine fields, in-field Mössbauer spectra clearly showed two different $Fe^{3+}$ sites, demonstrating that the sample is ferrimagnetically ordered. The two spinel sites are found to behave differently under an external field of 12 T: whereas the moments located at A sites show a perfect alignment with the external field, spins at B sites are canted by an angle $\alpha_B = 49\pm2$ °. We discuss the significance of this particle structure for the observed magnetic behavior.

**Short Title:** Ferrimagnetism and spin canting of $Zn(^{57}Fe)_2O_4$ nanoparticles.

PACS: 75.50.Pp, 76.80.+y , 81.05.Ys , 75.50.Gg

## INTRODUCTION

Magnetic nanoparticles dispersed in a nonmagnetic matrix have been the focus of an increasing amount of research work for the last decade [1-4]. Both theoretical models [5,6] and many potential technological applications of nanostructured devices [7-9] need a mindful characterization of the interplay between particle size distribution, dispersion and magnetic interactions. Different strategies have been employed to control particle distributions, including synthesis in gas phase, sol-gel chemistry, polymer blends, solvent chemistry and high energy milling for obtaining fine dispersions and ferrofluids.[10-16]

Zinc ferrite ($ZnFe_2O_4$) has been a particular subject of study because of its contrasting magnetic properties compared to other spinel ferrites, e.g. low ordering temperature and antiferromagnetic ground state. This particular compound has a normal spinel structure, in which the iron atoms are located at B (octahedral) sites and Zn atoms occupy the tetrahedral A sites. Whereas most iron-rich ferrites are ferrimagnetically ordered at room temperature, the antiferromagnetic transition temperature ($T_N$) of stoichiometric, well-crystallized $ZnFe_2O_4$ is $T_N = 10\pm1$ K. [17,18] This difference is originated in the much weaker superexchange interactions between B sites than the corresponding A-B interactions. For this reason, a small migration of Fe atoms from B to A sites can originate ferrimagnetic regions with strong A-B superexchange interactions. In fact, this partial inversion has been frequently invoked as the source of the much larger $T_N$ observed in $ZnFe_2O_4$ fine particles [19,20], but it remains unclear whether the observed inversion is due to the chemical routes usually employed for nanoparticle synthesis, such as sol-gel and mechanical grinding, or is an intrinsic property of nanometric $ZnFe_2O_4$ particles due to finite-size effects.

In this work, we present a structural and magnetic characterization of nanocrystalline $ZnFe_2O_4$ spinel particles grown within a ZnO matrix. Aimed to produce nanostructured systems composed of *well crystallized* magnetic particles, we have taken advantage of the well-known fact that trivalent iron has an extremely low solubility on ZnO phase [21], and chemical kinetics of the Fe-Zn-O system favors the formation of the more stable $ZnFe_2O_4$ phase instead of any of the iron oxides. We present evidence that the expected SPM behavior of such a system is related to the ferrimagnetic order of A and B sublattices, which have very different local anisotropies and display localized canting of the magnetic moments at the B sites.

## EXPERIMENTAL

The samples studied in this work were prepared by dissolving zinc oxide (99.99 %) and metallic $^{57}Fe$ (enriched to ~ 98 %) in a 0.2 M solution of $HNO_3$, to obtain a nominal composition $Zn_{0.99}Fe_{0.01}O$. The precipitate was gently warmed at 100 °C until complete dried, and afterwards fired in air at 500 °C, 750 °C and 950 °C for 24 h with several intermediate grindings. After the last annealing, samples were slowly cooled (1 K/min) to room temperature. Structural data was obtained from x-ray diffraction patterns collected in a commercial diffractometer with Cu-K$\alpha$ radiation, and the resulting patterns were refined by Rietveld profile analysis [22,23]. High Resolution Transmission Electron Microscopy (HRTEM) investigations were performed using a 200 kV Model CM200 Philips electron microscope, conditioning the samples by dropping an alcohol-powder suspension on a carbon-coated nickel grid. Mössbauer spectra were recorded in transmission geometry between 4.2 K and 300 K using a 50 mCi $^{57}Co$/Rh-matrix source in constant acceleration mode. An external magnetic field up to 12 T was applied parallel to the direction of the $\gamma$-ray using a superconducting coil. The spectra were fitted to Lorentzian line shapes using a non-linear least-squares program, calibrating the velocity scale with $\alpha$–Fe at room temperature. For relaxing spectra near the ordering temperature, the fits were obtained using static hyperfine field distributions, trough non-linear least-squares algorithms. To perform magnetic measurements, the powder was dispersed in epoxy resin and molded in cylindrical shape (5mm diameter x 3mm height). Magnetic measurements were done using a commercial SQUID magnetometer in both zero-field-cooling (ZFC) and field-cooling (FC) modes, between $1.8\ K \leq T \leq 400\ K$ and applied fields up to 70 kOe. Data were corrected for diamagnetic contribution of the sample holder and for core diamagnetism of the ions.

## *RESULTS AND DISCUSSION*

At room temperature, x-ray patterns were indexed with two phases (Fig. 1): the ZnO (zincite) hexagonal structure, space group $P6_3mc$ with cell parameters a = b = 3.250(1) Å and c = 5.210(1) Å, and small peaks corresponding to the most intense reflections of cubic (space group Fd3m) $ZnFe_2O_4$ spinel phase (franklinite). Detailed analysis of these lines showed that the full width at half-maximum (FWHM) of the three major peaks of the spinel phase, corrected for instrumental broadening was 0.18±0.04°, much larger than the corresponding ZnO peaks (FWHM = 0.06±0.01°). Assuming that strain effects do not contribute to the linewidth, an average crystallite size <D> = 23.4±0.9 nm was estimated using Scherrer equation. This indicates that the $ZnFe_2O_4$ phase consists of nanometric particles dispersed in the ZnO matrix. When included in the refinement process, the relative amount of $ZnFe_2O_4$ phase estimated from the profile analysis was 1.1±0.8%, which is in the limit of x-ray detection. The refined cell parameter for the spinel phase was a = 8.45(1) Å. The bright field TEM micrographs, together with an energy dispersive x-ray spectrometer (EDS) confirmed that the $ZnFe_2O_4$ nanoparticles are rather uniformly dispersed at the grain boundaries of the crystalline ZnO matrix. The particle morphology, on the other hand, is found to vary from nearly spherical to stick-shaped. These two extreme cases can be seen in figure 2, where the selected area image shows two nearly spherical particles of ~100 nm (light-gray spots on the figure), on which several rounded- and stick-shaped $ZnFe_2O_4$ particles are clearly observed (black spots on the border of the ZnO particles). From the inspection of several TEM areas we observed that the morphology of the grains were between these two extreme shapes and 20-30 nm in size, in good agreement with the sizes inferred from XRD data. The selected area diffraction (SAD) pattern on the black spots ($ZnFe_2O_4$ particles) clearly demonstrated a good crystallinity of the spinel phase.

Zero-field cooling (ZFC) and field cooling (FC) magnetization curves taken at H = 1 kOe (fig. 3) show a maximum in both ZFC and FC curves at a temperature $T_M \sim 17$ K that depends on the applied field, and below which irreversible behavior sets in. As shown in the inset of figure 3, for high temperatures the magnetization follows approximately a Curie-Weiss (CW) law, and thus the data was fitted in this range (defined as T > 220 K) using the expression $\chi_{DC} = C/(T-\Theta)$. From this fit an effective magnetic moment of $\mu_{eff}$ = 38(1) $\mu_B$/(f.u. $ZnFe_2O_4$) was obtained, a much larger value than the corresponding for paramagnetic single $Fe^{3+}$ atoms ($\mu_{eff} = 5.9\mu_B$), indicating the existence of SPM clusters as the magnetic entities at high temperature. Magnetization vs. field curves at different temperatures (fig.4) show that magnetic interactions are still present well above $T_M$, suggesting a modified cation distribution respect to the normal (AFM) spinel configuration. It can be also noted that at these high temperatures the system does not saturate even for applied fields of 7 Tesla. By extrapolating the high-field region of the $M(H^{-1})$ curve taken at T = 200 K, we estimated the saturation magnetization $M_S$(T=200K) = 0.028(3) $\mu_B$/f.u. $ZnFe_2O_4$. The measured coercive field $H_C = 25\pm10$ Oe remains essentially constant down to $T = T_M$, and increases steeply for $T < T_M$ attaining a value of $H_C = 670\pm10$ Oe at 2K (see the inset of Fig. 4). Additionally, it can be observed in figure 5 that there is no superposition of M(H/T) at different temperatures *above* the transition, as expected for a system of non-interacting SPM particles. The $M_S$ value at T = 2K estimated from the high-field region was $M_S(T=2K) = 0.134(7)$ $\mu_B$/f.u. $ZnFe_2O_4$.

As the Fe ions migrate to A sites, the stronger A-B superexchange interactions yield ferrimagnetic ordering of iron atoms. The low $\mu_{eff}$ value observed at both high and low temperatures could be explained within two rather different scenarios. In the first, the resulting $ZnFe_2O_4$ particles have a normal Fe distribution (i.e., all Fe ions at B sites) but the AFM state resulting from the weak $J_B$-$J_B$ superexchange interactions has not a collinear but a canted structure of the magnetic moments. From the preparation method used in the present samples, involving a last step with slow cooling to room temperature, the spinel phase is formed within the matrix in equilibrium conditions, and thus a similar situation than bulk $ZnFe_2O_4$ (i.e. normal configuration) could be expected. However, as the low ordering temperature of bulk $ZnFe_2O_4$ (i.e. $T_N$ = 10(1) K) is a result of the weak $J_B$-$J_B$ exchange interactions, the hypothesis of a normal configuration does not account for the magnetic properties observed in M(H) curves at room temperature, since B-B interactions are not operative for T > 10 K and thus the samples should display paramagnetic behavior (i.e. linear dependence of the M *vs.* H curves).

In the second possible scenario, the $ZnFe_2O_4$ particles have a nearly inverted structure, in which only a negligible fraction of the Fe spins are uncompensated. In this situation, the stronger $J_A$-$J_B$ exchange interactions could result in magnetically ordered particles even above room temperature, thus yielding the observed SPM-like behavior. This hypothesis implies that the observed ordering process at $T_M$ does not correspond to the transition *within* the particles, but to a collective, interparticle ordering process. This is in agreement with previous a.c. susceptibility measurements on these samples [24] that showed a strong dependence of the ordering temperature $T_M$ with frequency. Although the precise nature of the transition observed at $T_M$ is not yet clear, this frequency dependence points to some SPM blocking or spin-glass-like transition from particle interactions, and discards the possibility of an ordinary para- to ferromagnetic (or AFM) transition at this temperature.

Mössbauer spectra at room temperature (see Fig. 6 and Table I) were fitted with a doublet whose hyperfine parameters were coincident to previous data on $ZnFe_2O_4$ nanoparticles. [18,19] Unfortunately, these parameters are essentially the same that the expected for the bulk material in the paramagnetic state[21,25], making impossible to assert whether this doublet correspond to a paramagnetic or a superparamagnetic state of the $ZnFe_2O_4$ phase. No other secondary Fe-containing phases were detected within ≈1% of experimental accuracy. The hyperfine parameters of the doublet remains essentially unchanged for lower temperatures down to T ~30 K, where some relaxation effects begin to develop. At T = 4.2 K the spectra show that the sample is magnetically ordered, indicating a transition to a frozen state at some temperature in between. The zero-field Mössbauer spectra below the ordering temperature was composed of six broad absorption lines, which could be fitted with two hyperfine sites. However, the difference of the hyperfine fields of these two sites (~3.8 T, see Table I) is too small to perform confident fits due to the strong overlap of the two six-line patterns. This overlapping is known to produce misleading interpretations in some spinel compounds like $CuFe_2O_4$, where unphysical results are obtained for the iron population if the spectral areas are considered.[26] Therefore, the iron populations at A and B sublattices will be analyzed concurrently with the discussion of in-field Mössbauer data (see below).

Figure 7 shows the evolution of the Mössbauer spectra when crossing the ordering temperature, where a complex shape due to the relaxation of the iron magnetic moments can be noticed. To get a qualitative picture of the ordering process we have fitted the spectra for T ~ $T_B$ with a static hyperfine field distribution at each temperature (see figure 7). We mention here that, for such transitions involving particle size (and the corresponding relaxation time) distributions, a dynamical model for relaxation time distributions accounting for the evolution of the lineshape should be used instead a static one. Unfortunately, for the present case of two different hyperfine sites such a model has not yet been developed. Therefore, the results from the static distribution fitting should be interpreted as approximate and only qualitative information is to be extracted. As can be seen from figure 8, the average hyperfine field $B_{hf}(T)$ (calculated from the static distribution) displays a sudden drop to $B_{hf} = 0$ at some point between 20 and 25 K, in agreement with the observed ordering temperature from magnetization data.

By comparing spectra taken at 4.2 and 10 K, thermally induced spin fluctuation are readily observable through the reduction of $B_{hf}$, suggesting that even at 4.2 K these collective excitations are still operative and system is not fully ordered. As the temperature is increased, the fluctuations of the magnetic spin direction through 180° become dominant and for $T \geq 15$ K magnetic relaxation sets in. In order to suppress thermal fluctuations already observed at 4.2 K, and to better resolve the overlapping sextets, we have undertaken Mössbauer measurements in 12-Tesla applied field at T = 4.2 K. Figure 9 shows that under the applied field, two distinct spectral components can be readily resolved. The isomer shift values for both subespectra correspond to $Fe^{3+}$ oxidation state. However, these two signals have rather different effective hyperfine magnetic fields, as could be expected for the dominant $J_{A\text{-}A}$ and $J_{B\text{-}B}$ superexchange interactions (as compared to $J_{A\text{-}B}$) that originate a ferrimagnetic alignment of magnetic moments. When an external field $B_{ext}$ is applied (in longitudinal geometry, as in the present setup), it will add to the internal hyperfine fields $B_{int}$, yielding effective fields ($B_{eff}$) larger (lower) when the spins are parallel (antiparallel) to the direction of $B_{ext}$. Additionally, it can be noticed that the $\Delta m = 0$ lines (i.e., the second and fifth lines) have either small or zero intensities for the outermost sextet, whereas the sextet with smaller hyperfine field has lines 2 and 5 well defined.

The hyperfine parameters from the least-squares fit are displayed in Table I. The assignment of each subespectra to A and B sites was mainly based on their relative population. The patterns amounting 19(4) % and 81(4) % of the total resonant area were assigned to $Fe^{3+}$ at A and B sites respectively, since iron is known to occupy mainly octahedral B sites. The A-site population is very similar to previous results on $ZnFe_2O_4$ nanoparticles obtained by the supercritical sol-gel method [27] and ball-milling [28]. However, a remarkable difference between those works and the present in-field Mössbauer data is related to the existence of *only two magnetically ordered subespectra.* This is a clear evidence of ferrimagnetic order of our particles, which is related to the strong A-B superexchange interactions in partially inverted ferrites. The signal associated to the octahedral B sites show essentially zero QS values as should be expected due to their cubic symmetry, whereas the tetrahedral A sites have small but definite QS values, which increase with the applied field. The outer peaks of the spectral component corresponding to the A site split to higher velocities, implying that A spins are antiparallel to the applied field, whereas the B spins are parallel (the peaks shift to lower velocities).

Another striking feature of the present data is the observed difference in the ratio between the relative spectral areas $A_i$, $q = A_2/A_1$ of lines 1 and 2 (or $A_5/A_6$) which is a direct indication of spin canting. An estimation of the mean spin canting angle can be obtained from the relative intensities by the relation

$$\sin^2 \alpha = \frac{6q}{4+3q} \qquad (1)$$

where $\alpha$ is the angle between the magnetic quantisation axis (assuming that the magnetic interaction is dominant) and the direction of the gamma-ray beam (i.e., the applied field $B_{app}$ in the present experimental setup). From this relationship we see that q = 0 implies $\alpha$ = 0 meaning that the spins are parallel to the applied field, and for q = 4/3 we have $\alpha = 90°$.

Figure 9 and Table I clearly show that the spin configuration inferred for the A sublattice is nearly parallel to the external field (i.e. $q_A = 0.01$, $\alpha_A = 180$ °), whereas the magnetic moments at B sites are canted an angle $\alpha_B = 49\pm2$ °. This result is also in contrast to previous works where canting in both A and B sublattices was observed.[27,28] The selective canting of magnetic moments at B sublattice implies very different local anisotropies for A and B sites.

A second method for calculating the canting angles are based on the vector sum (fig. 10) of the local (hyperfine) and applied fields

$$B^2_{hyp}(i) \ = \ B^2_0(i) + B^2_{app}(i) + B_{app}(i)\, B_0(i) \cos\alpha \qquad (2)$$

where $\alpha$ is the angle between the hyperfine field at the nucleus $B_0$ (in zero applied field) and the direction of the $\gamma$ ray (which is parallel to the applied field $B_{app}$ in the present setup). This second method is less precise than the intensity method, since for the present setup the samples are disk-shaped and placed perpendicular to the applied field, thus maximizing the effects of the demagnetizing field. By considering this source of uncertainty to be ~2 T (from measurements of the demagnetizing field on the calibration $\alpha$-Fe foil with the same geometry), we have estimated values $\alpha_A = 0\pm10$ ° and $\alpha_B = 58\pm10$ °, which are in reasonable agreement with the results obtained from Eq. 1.

From the above data it is inferred that there is a measurable inversion degree (~20 %) of the Fe ions, thus we are inclined to support the landscape of $ZnFe_2O_4$ particles with partially inverted structure. The presence of iron ions at both A and B sublattices is likely to originate strong coupling (*via* the $J_A$-$J_B$ exchange interactions) between both sublattices and thus high ordering temperatures. In these magnetically ordered particles, SPM behavior can be expected even at room temperature. Although this state is difficult to distinguish from a paramagnetic behavior, the observed frequency dependence of the transition temperature $T_M$, together with the absence of any other transition above $T_M$ support the picture of SPM behavior of the $ZnFe_2O_4$ particles above $T_M$.

In conclusion, we have carefully characterized the magnetic properties of crystalline $ZnFe_2O_4$ nanoparticles, grown within a nonmagnetic ZnO matrix. We have shown that the resulting particles have large population of Fe atoms at tetrahedral sites, yielding strong Fe(A)-Fe(B) superexchange interactions. These interactions in turn allow the magnetic nanoparticles to have an ordered spin structure resulting in SPM behavior even at room temperature. Furthermore these nanoparticles display some collective ordering at $T \sim 17$ K, whose precise nature is yet to be settled. A selective canting at B sites of the ferrimagnetic structure was inferred from in-field Mössbauer data, indicating different local anisotropies at each of these spinel sites.

## *ACKNOWLEDGEMENTS*

This work was supported by Brazilian agencies FAPESP and CNPq through Grants No. 01/02598-3 and 300569/00-9, respectively.

## *REFERENCES*

[1] Parker M A, Coffey K R, Howard J K, Tsang C H, Fontana R E and Hylton T L 1996 *IEEE Trans. Magn.* **32** 142.

[2] Tamari K, Doi T and Horiishi N 1993 *Appl. Phys. Lett.* **63** 3227.

[3] Lee J S, Kim K Y, Noh T H, Kang I K and Yoo Y C 1994 *IEEE Trans. Magn.* **30** 4845.

[4] Jonsson P, Svedlindh T and Hansen M F 1998 *Phys. Rev. Lett.* **81** 3976.

[5] Ziese M, Sena S, Shearwood C, Blythe H J, Gibbs M R J and Gehring G A 1998 Phys. Rev. B **57** 2963.

[6] Denisov S I and Trohidou K N 2001 Phys. Rev. B **64** 184433.

[7] Kim T W, Choo D C, Shim J H, and Kang S O 2002 Appl. Phys. Lett. **80** 2168.

[8] Mais N, Reithmaier J P, Forchel A, Kohls M, Spanhel L, and Muller G 1999 Appl. Phys. Lett. **75** 2005.

[9] Godovsky D Y 2000 Adv. Polym. Sci. **153** 163.

[10] Hasmonay E, Depeyrot J, Sousa M H, Tourinho F A, Bacri J C, Perzynski R, Raikher Y L and Rosenman I 2000 J. Appl. Phys **88** 6628.

[11] Choi M 2001 Nanopart J. Res. **3** 201.

[12] Muh E, Frey H, Klee J E and Mulhaupt R 2001 Adv. Funct. Mater **11** 425.

[13] Van Hyning D L, Klemperer W G and Zukoski C F 2001 Langmuir **17** 3128.

[14] Dubois E, Cabuil V, Boue F, and Perzynski R 1999 J. Chem. Phys. **111** 7147.

[15] Shinde S R *et al* 2000 J. Appl. Phys. **88** 1566.

[16] Zhou W L, Carpenter E E, Lin J, Kumbhar A, Sims J, and O'Connor C J 2001 European Phys.J. D **16** 289.

[17] Krûpica S and Novak P 1982 "*Oxide Spinels*", in *Ferromagnetic Materials*, edited by E. P. Wolfarth (North-Holland, Dordrecht), Vol. 3.

[18] Goya G F and Rechenberg H R 1999 J. Magn. Magn. Mater. **196-197** 191.

[19] Schiessl W *et al* 1996 Phys. Rev. B **53** 9143.

[20] Sepelak V *et al* 1998 J. Sol. St. Chem. **135** 52.

[21] Goya G F *et al* 1995 Solid State Commun. **96** 485.

[22] Rietveld H M 1969 J. Appl. Crystallogr. **2** 65.

[23] Skthivel A and Young R A 1991, "*DBWS-9006PC, Program for Rietveld analysis of X-ray powder diffraction patterns*", Georgia Institute of Technology, Atlanta.

[24] Goya G F 2002 IEEE Trans. Magn., **38** (September, 2002). In press.

[25] Mizoguchi T and Tanaka M 1963 J. Phys. Soc. Japan **18** 1301.

[26] Evans B J and Hafner S 1966 Phys. Lett. **23** 24.

[27] Oliver S A, Hamdeh H H and Ho J C 1999 Phys. Rev. B **60** 3400.

[28] Chinnasamy C N *et al* 2000 J. Phys. Cond. Mater. **12** 7795.

## *FIGURE CAPTIONS*

Figure 1: X-ray diffraction patterns of the $[Zn_{0.99}Fe_{0.01}]O$ system. The symbols indicate the positions of the indexed (*hkl*) reflections for the $ZnFe_2O_4$ (circles) and ZnO (down triangle) phases. The (*hkl*) numbers for each line are also shown.

Figure 2: Selected area TEM micrographs, showing the different $ZnFe_2O_4$ particle morphologies (dark spots) at the ZnO grain boundaries (pale areas).

Figure 3: Zero-field cooling (ZFC) and field cooling (FC) magnetization curves for $ZnFe_2O_4$ taken at H = 1 kOe, showing magnetic irreversibility for $T < T_M$.

Figure 4: Magnetization vs. field at several temperatures for $ZnFe_2O_4$. These curves were obtained after field cooling in a 7 T field. Inset: Enlargement of the low-field region (up to 1 kOe) showing the increase of $H_C$ for T < 50 K.

Figure 5: Normalized magnetization vs. H/T taken at different temperatures. Note the lack of superposition of M(H/T) curves *above* the transition ($T > T_f$ ). For comparison, one curve at $T < T_f$ (T = 5 K) is shown.

Figure 6: Room temperature Mössbauer spectrum for Zn-Fe-O sample.

Figure 7: Mössbauer spectra of Zn-Fe-O sample at different temperatures crossing the ordering temperature $T_M$. The corresponding hyperfine field distributions at each temperature is shown on the right side.

Figure 8: Average hyperfine field $B_{hf}$ calculated from the static distribution for 4.2 K $\leq$ T $\leq$ 50 K.. The arrow indicates the zero value somewhere between 20 and 25 K.

Figure 9: Mössbauer spectrum taken in applied field $B_{app}$ = 12 T, at T = 4.2 K. The two components used to fit the spectrum are also shown.

Figure 10: Vector diagram of the hyperfine fields at A and B sites under the action of an external field. All vectors were drawn using an approximate scale.

***Table Caption:*** Hyperfine field (B), Isomer shift (IS) quadrupole splitting (QS), and linewidth (Γ), area ratios of lines $A_2/A_1$ (q) and spectral area (A) of each spectral component in $ZnFe_2O_4$ phase. Errors are quoted between parenthesis.

FIGURE 1

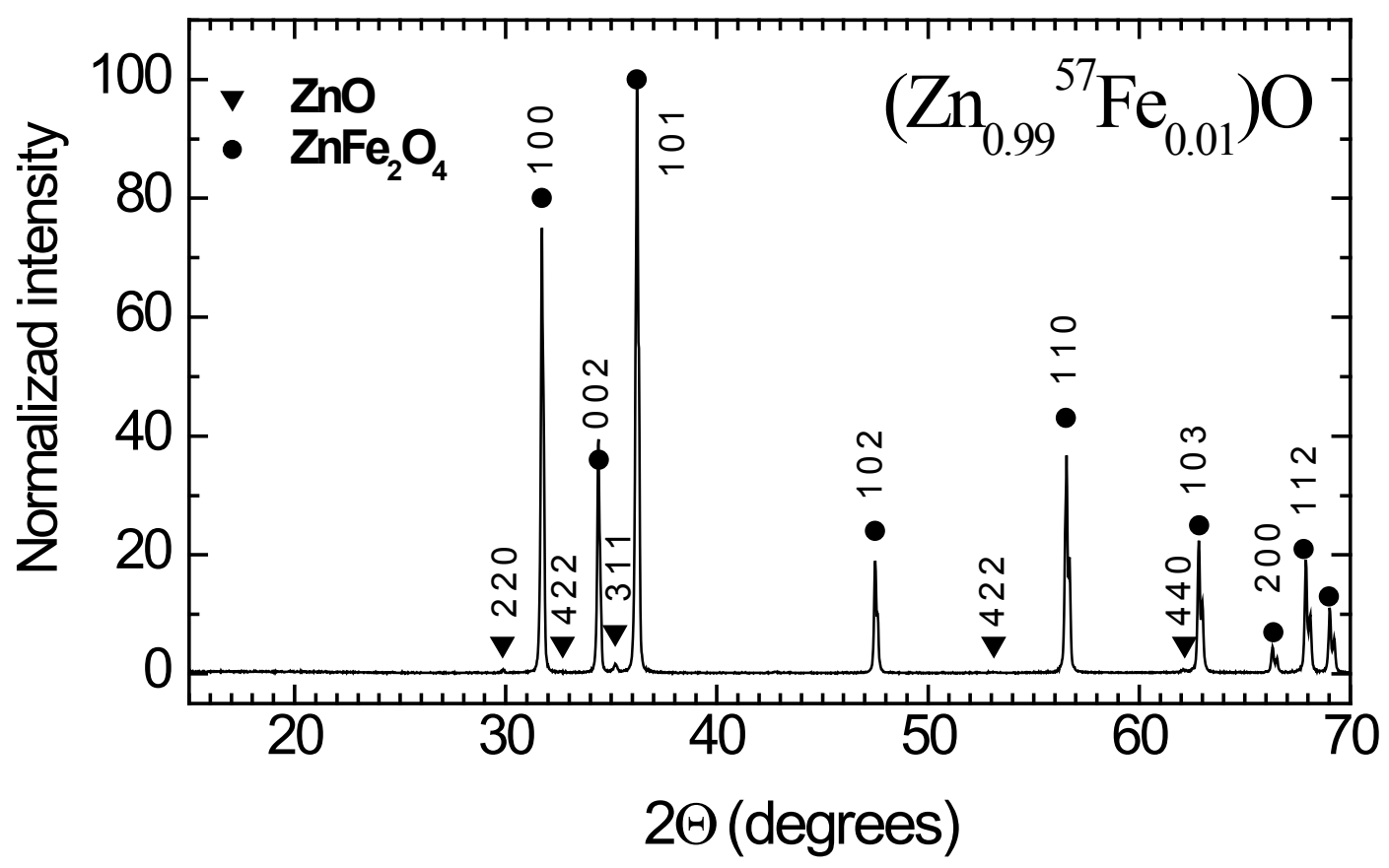

FIGURE 2.

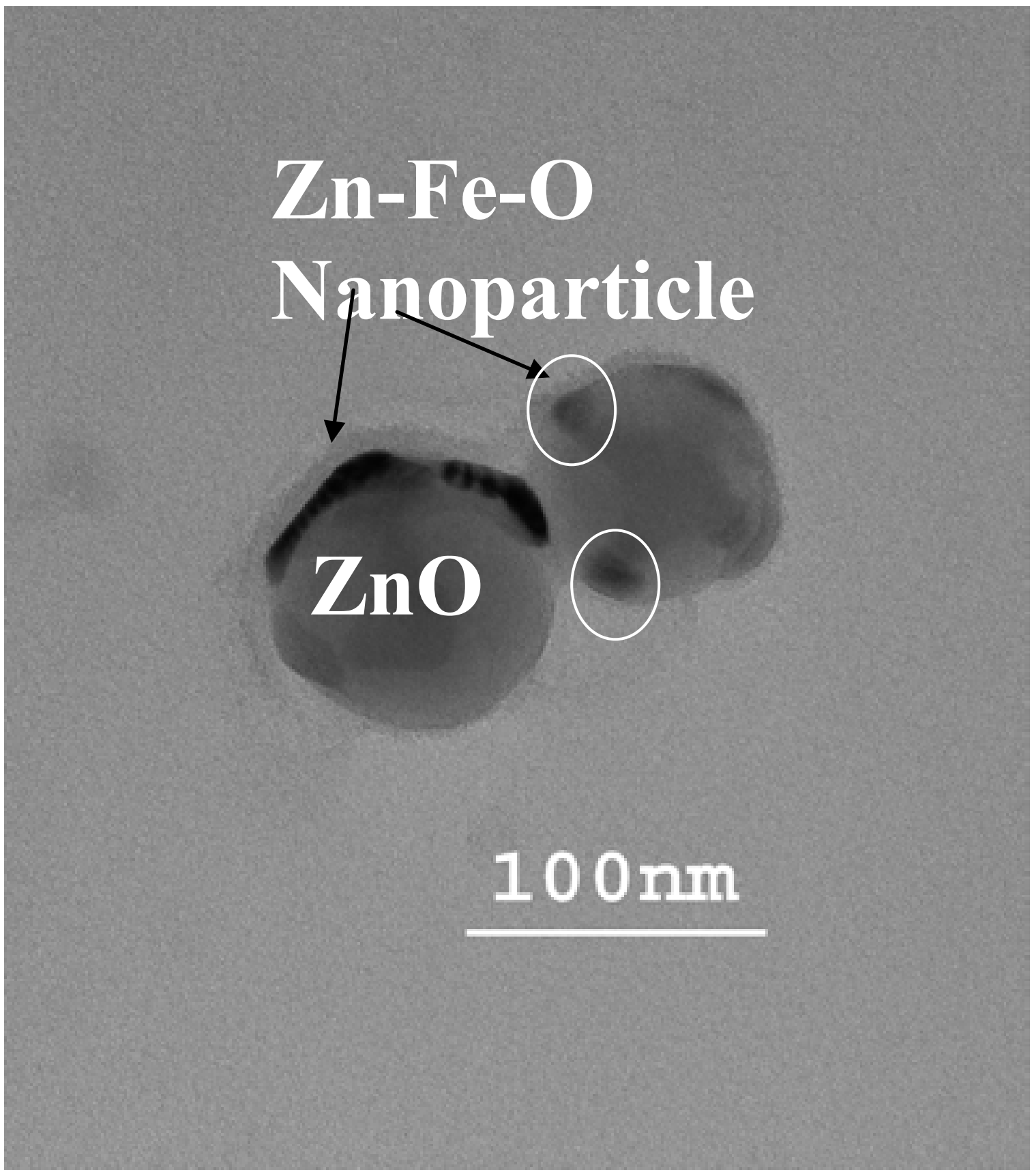

FIGURE 33

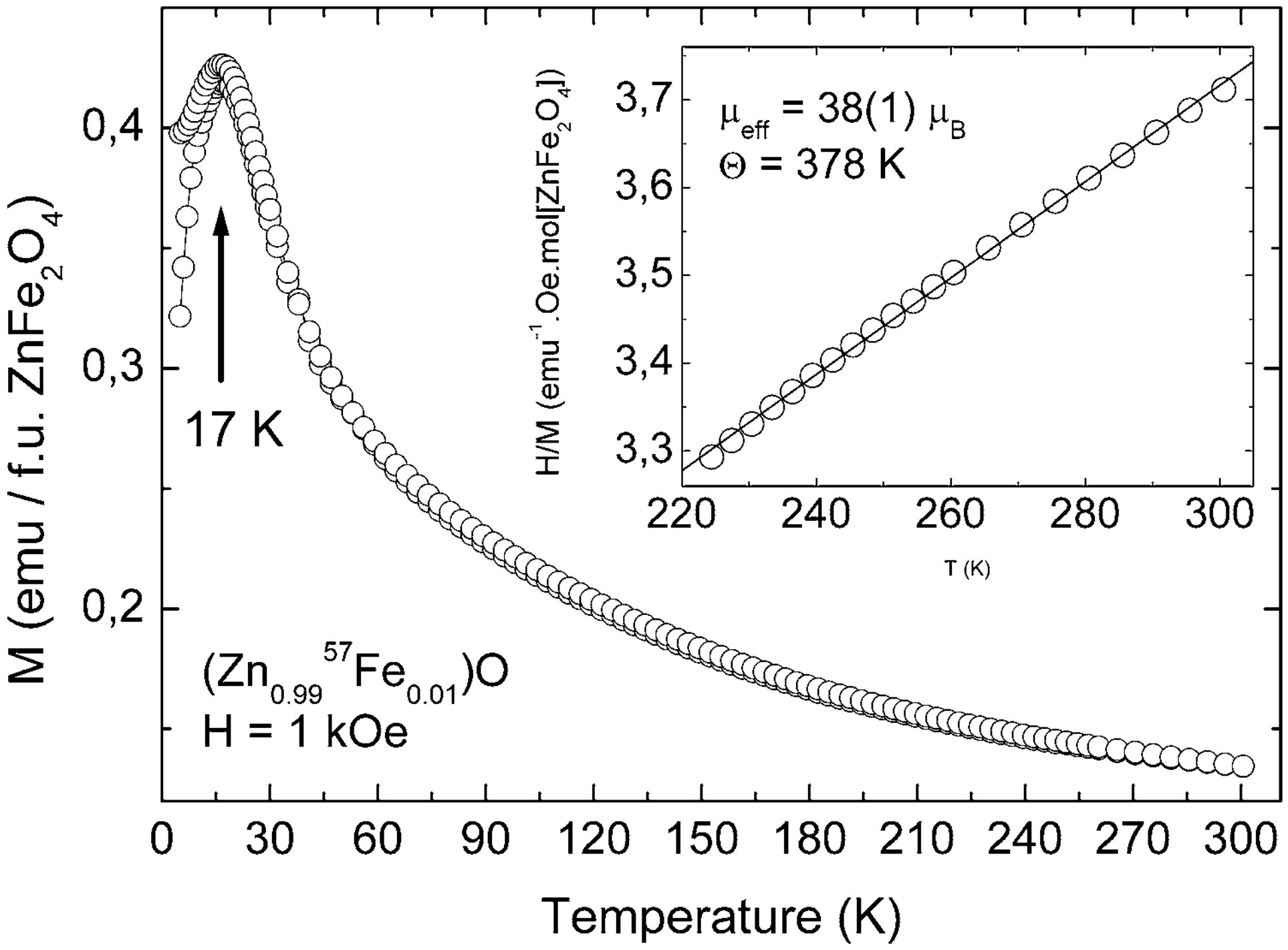

FIGURE 4

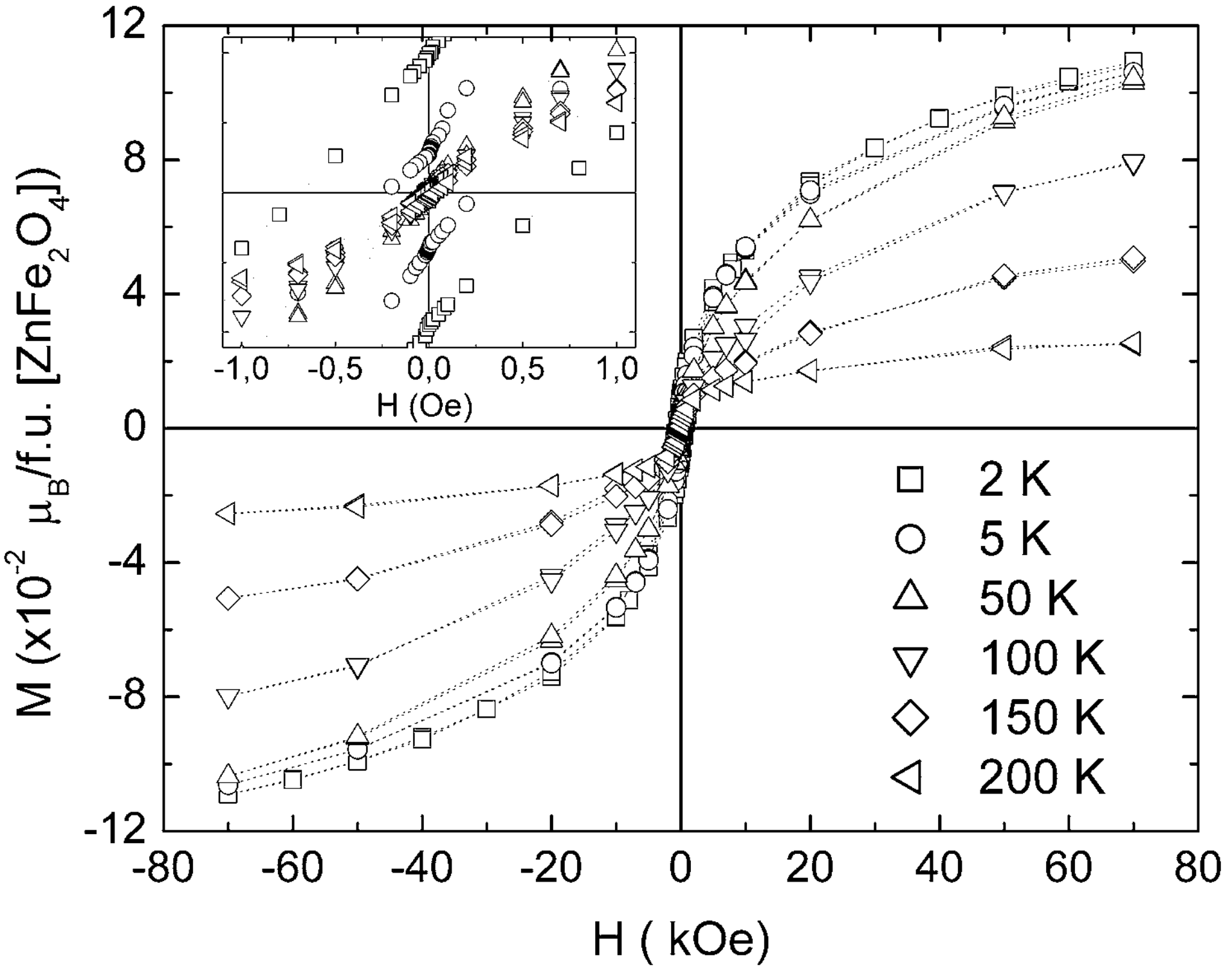

FIGURE 5.

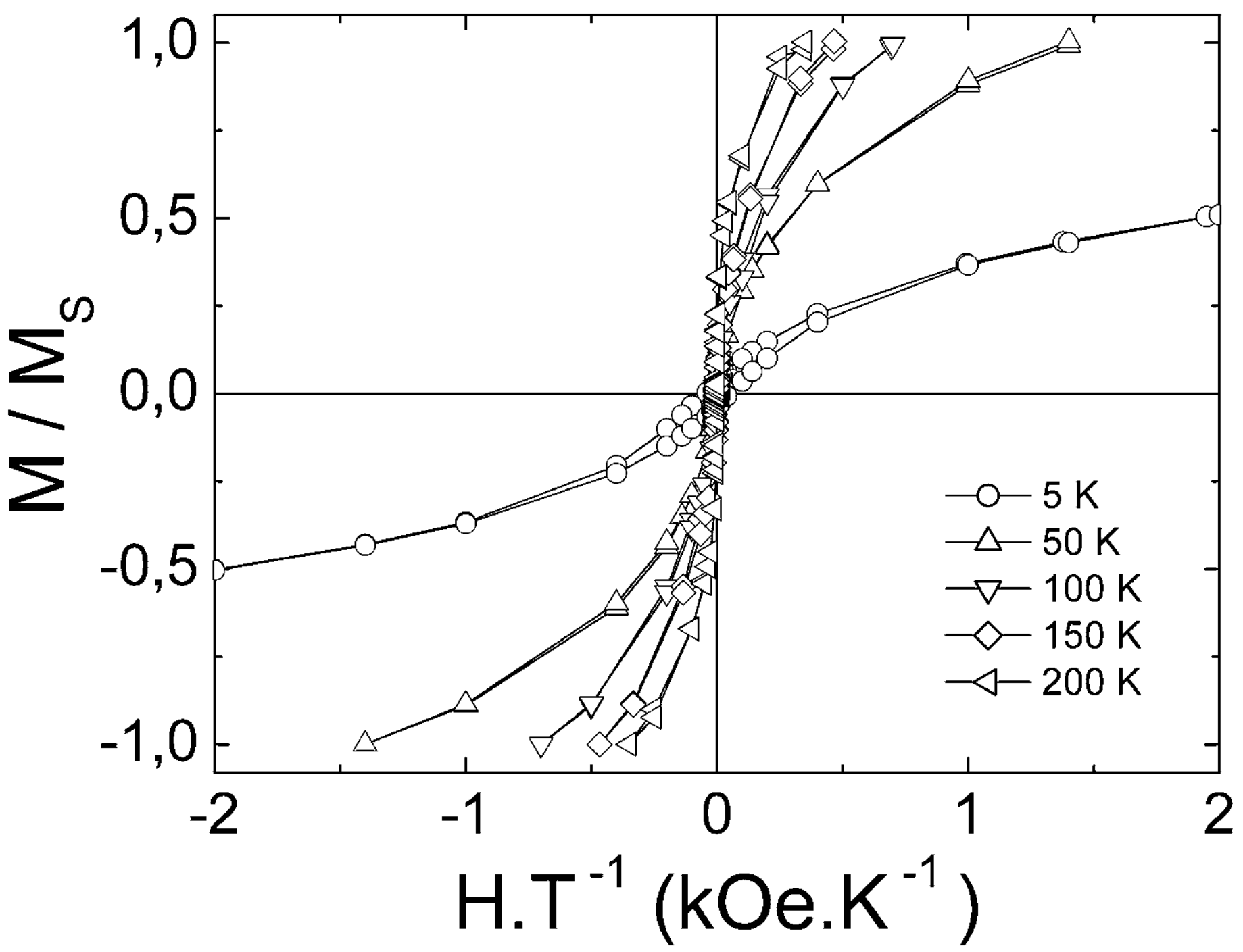

FIGURE 6.

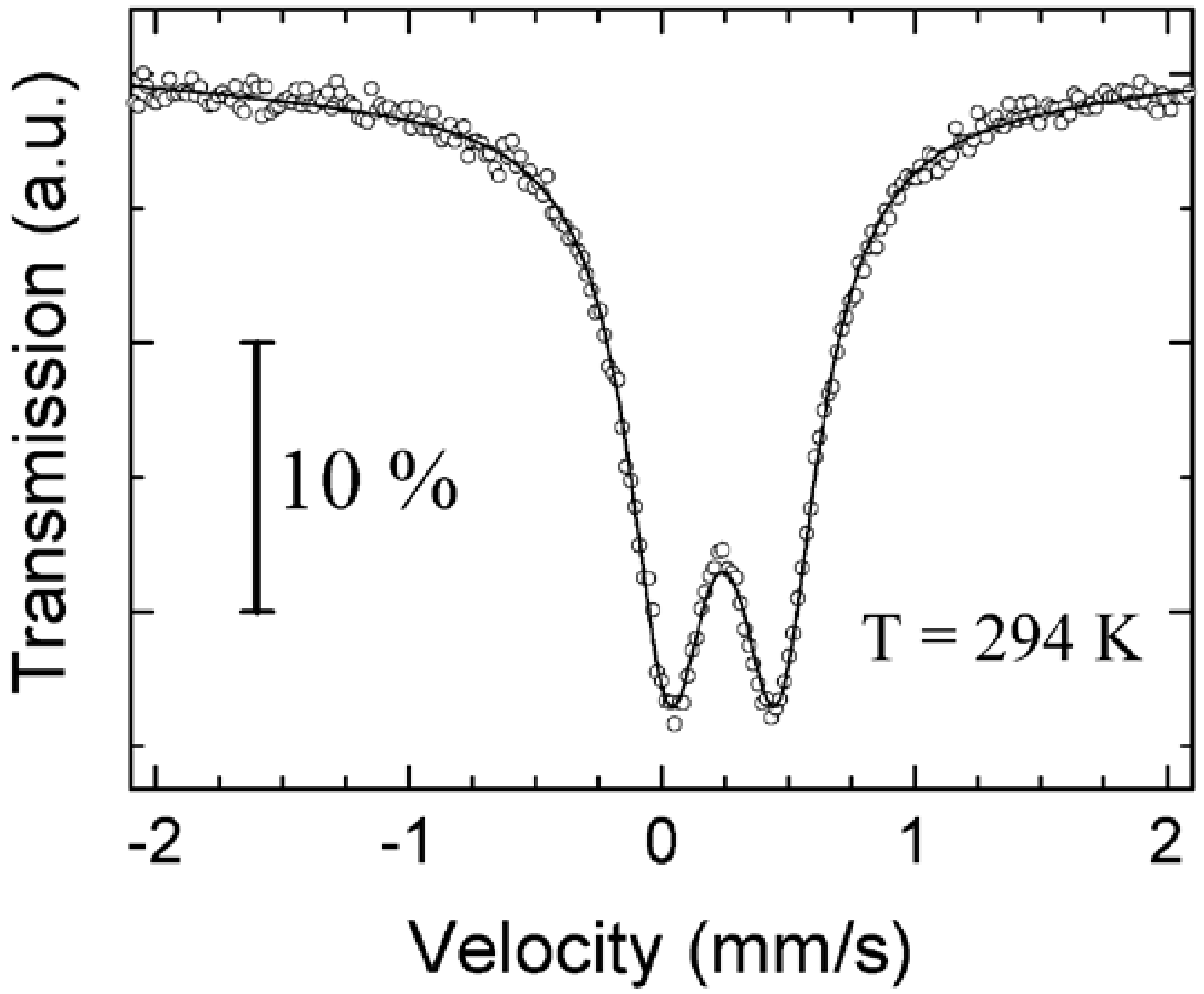

FIGURE 7.

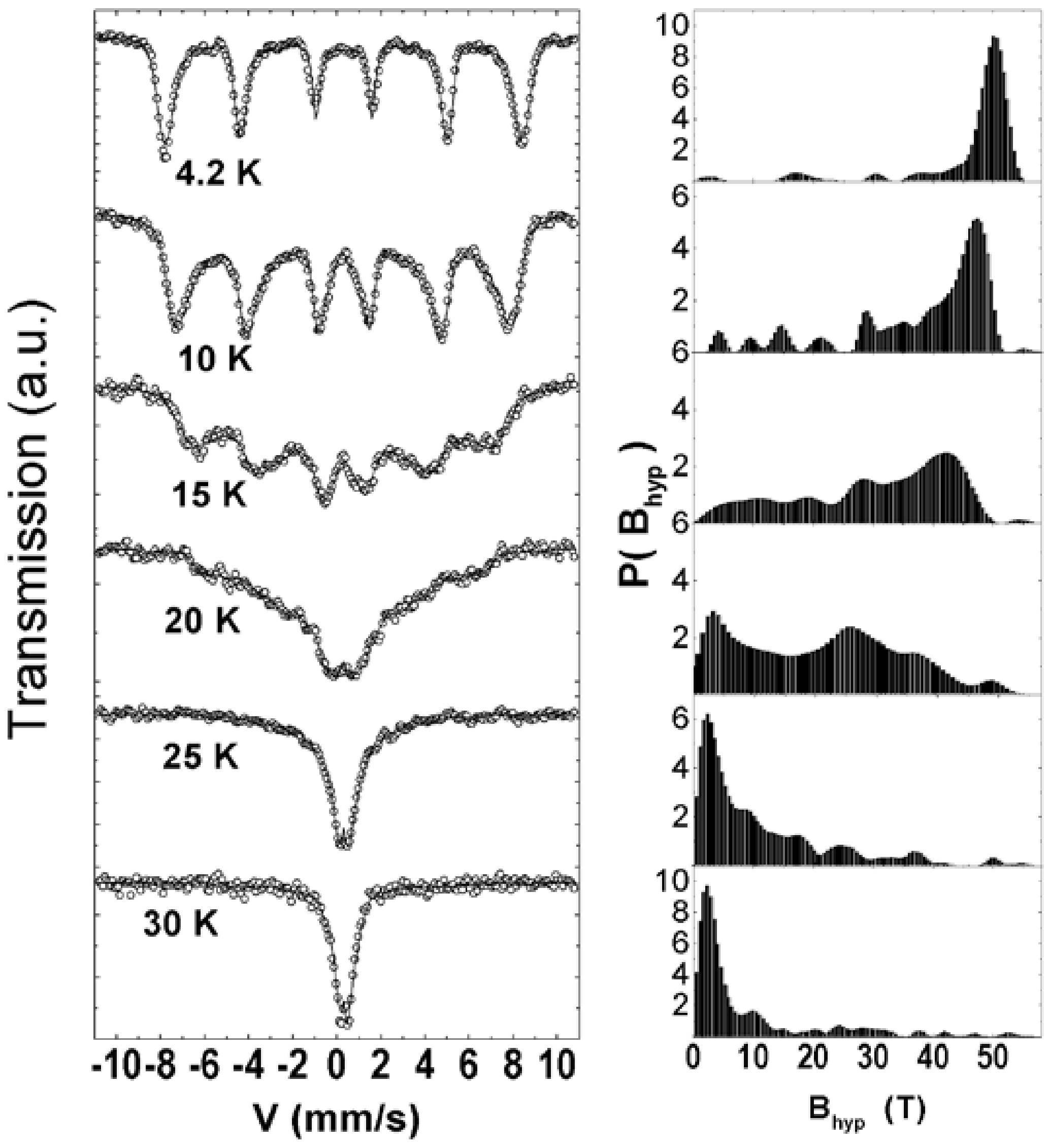
4.2 K
10 K
15 K
20 K
25 K
30 K
Transmission (a.u.)
-10 -8 -6 -4 -2 0 2 4 6 8 10
V (mm/s)
P( $B_{hyp}$ )
0 10 20 30 40 50
$B_{hyp}$ (T)

FIGURE 8.

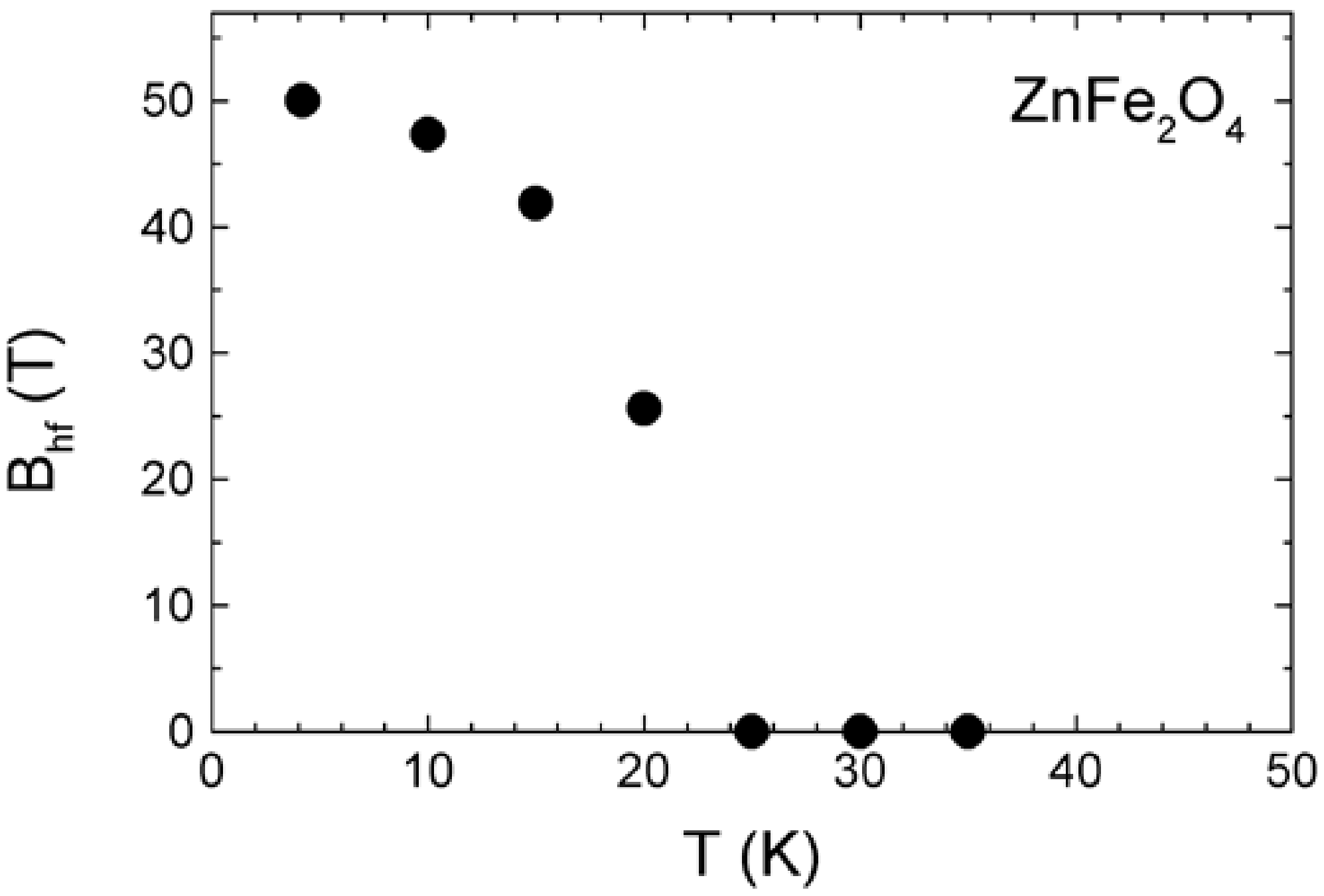

FIGURE 9.

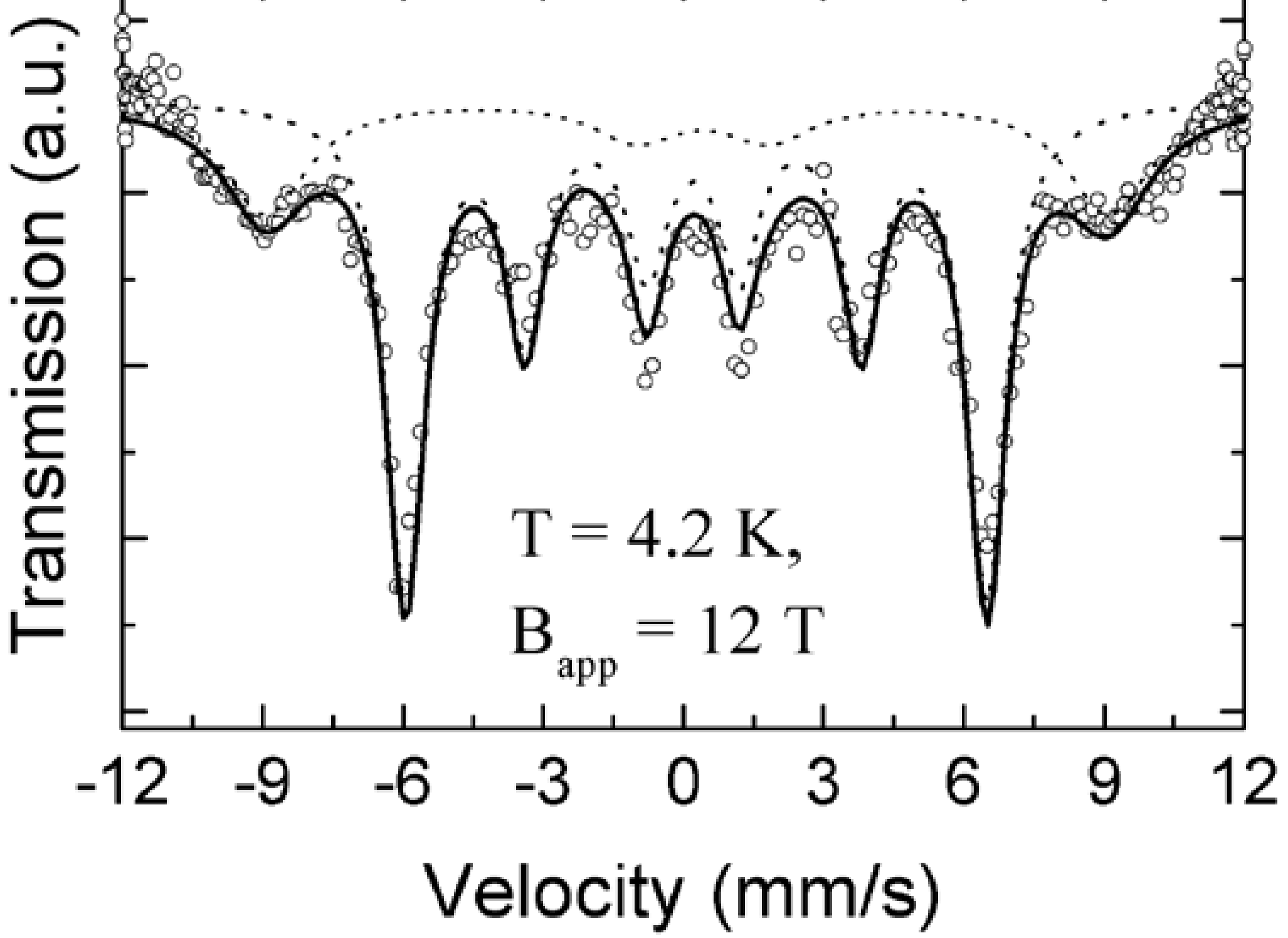

FIGURE 10.

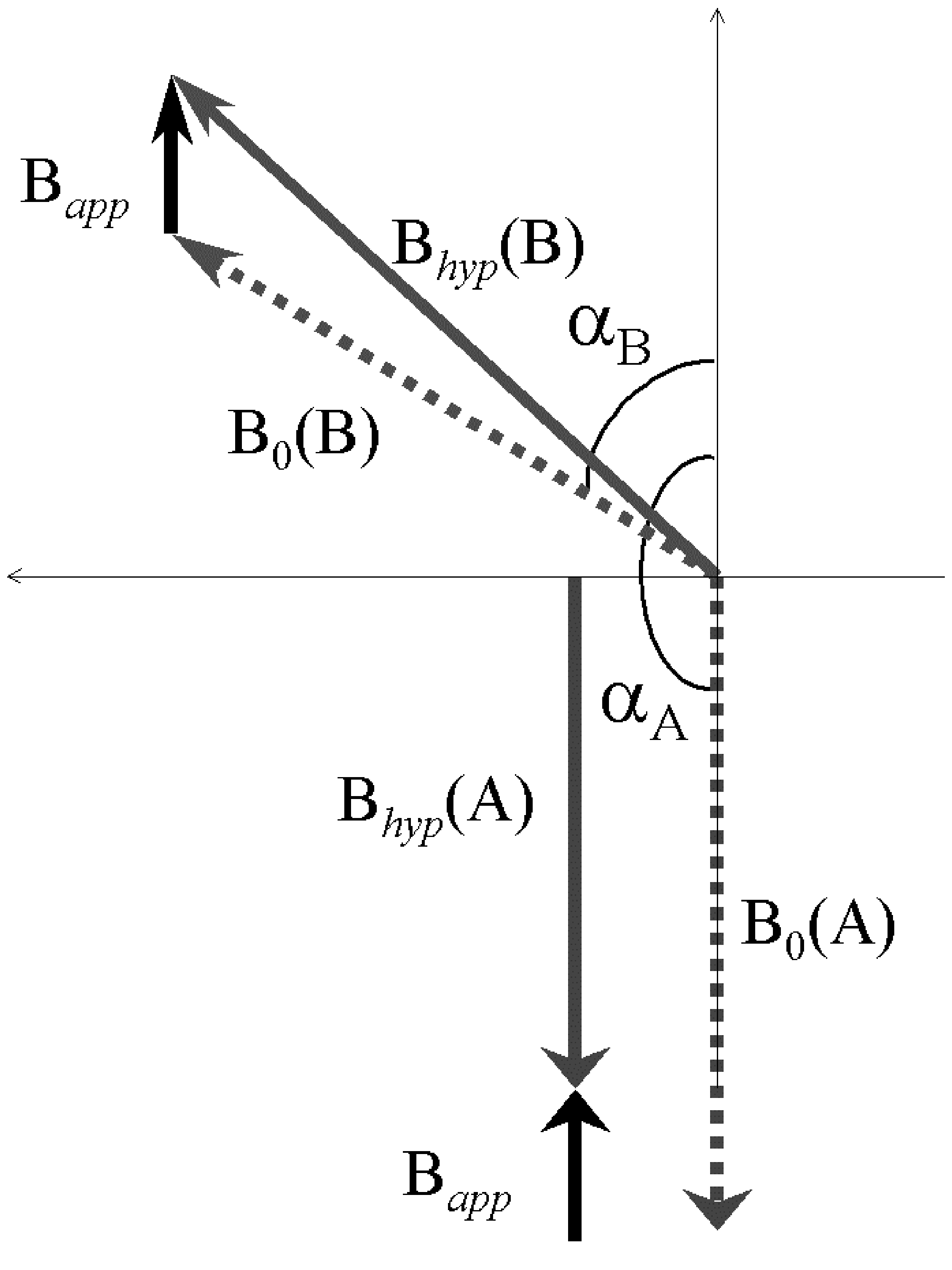

TABLE I.

| T | Site | B (Tesla) | IS (mm/s) | QS (mm/s) | Γ (mm/s) | q =$A_2/A_1$ | A (%) |
|---|---|---|---|---|---|---|---|
| 300 K | | -- | 0.35(1) | 0.43(1) | 0.39(1) | 1 | 100(2) |
| 4.2 K | A | 48.7(1) | 0.45(1) | 0.07(1) | 0.50(2) | 0.667 | 54(4) |
| | B | 50.9(1) | 0.44(1) | 0.00(1) | 0.40(2) | 0.667 | 46(4) |
| 4.2 K $B_{app}$=12T | A | 58.9(3) | 0.24(2) | 0.17(2) | 2.23(2) | 0.01 | 19(4) |
| | B | 38.6(2) | 0.24(2) | 0.03(2) | 0.77(2) | 0.55 | 81(4) |